\documentclass[letterpaper, 10 pt, conference]{ieeeconf}   
\IEEEoverridecommandlockouts                              
                                                          
\usepackage{subcaption}
\usepackage{color}

\usepackage{graphicx}
\newcommand{\rref}[1]{(\ref{#1})}

 \usepackage{amsfonts,amsmath,amssymb,color,setspace}

 \usepackage{graphicx,color}

\usepackage{comment}

\IEEEoverridecommandlockouts

\usepackage{color}

\newcommand{\ashalf}{\renewcommand{\arraystretch}{1.25}}
\newcommand{\ashalfplus}{\renewcommand{\arraystretch}{1.32}}

\newcommand{\mm}[1]{}
 \usepackage{amsfonts,amssymb,color,setspace}

 \usepackage{graphicx,color}

\title{Reference Governor Design in the Presence\\of Uncertain Polynomial Constraints}
\author{\hspace{1em} Rick Schieni, Chengwei Zhao, Michael Malisoff, and Laurent Burlion
\thanks{The first author  was supported by  Middlesex College. This research is supported in part by the Office of Naval Research (ONR) under award N00014-22-1-2135.}
\thanks{Rick Schieni is a graduate student with the Department of Mechanical and Aerospace Engineering, Rutgers University, Piscataway, NJ, 08854, USA (Rick.Schieni@Rutgers.edu)}
\thanks{Chengwei Zhao is a graduate student with the Department of Mechanical
and Aerospace Engineering, Rutgers University, Piscataway, NJ, 08854, USA (Chengwei.Zhao@Rutgers.edu)}
\thanks{Michael Malisoff is with the Department of Mathematics, Louisiana State University, Baton Rouge, LA, 70803, USA (Malisoff@lsu.edu)}
\thanks{Laurent Burlion is with the Department of Mechanical and Aerospace Engineering, Rutgers University, Piscataway, NJ, 08854, USA (Laurent.Burlion@Rutgers.edu)}
}
\begin{document}
\maketitle
\thispagestyle{empty}
\begin{abstract}
Reference governors are add-on schemes that are used to modify trajectories to prevent  controlled dynamical systems from violating input or state constraints, and so are playing  an increasingly important role in aerospace, robotic, and other applications.
Here 
we present a novel reference governor design 
for systems whose polynomial constraints depend on unknown bounded parameters. This is a significant departure from earlier treatments of reference governors, where the constraints were linear or known, because here we transfer the uncertainties into the constraints instead of having them in the  closed loop dynamics, which greatly simplifies the task of determining future evolution of the constraints.
Unlike our earlier treatment of reference governors with polynomial constraints which transformed the constraints into linear ones that depend on an augmented state of the system, here we transform the constraints into linear ones  
 that depend on both the system's state and uncertain parameters. Convexity allows us to estimate the  maximal output admissible set for an uncertain pre-stabilized linear system. We show that it is sufficient to only consider extreme values of the uncertain parameters when computing and propagating the polynomial constraints. 
 We illustrate our method using an uncertain longitudinal dynamics for civilian aircraft which is controlled using a disturbance compensation method and needs to satisfy   input and state constraints, and where our reference governor method ensures that safety constraints are always satisfied.
\end{abstract}
\section{Introduction} 
\label{sec:introduction}
Reference governors are playing an increasingly important role in current research in control engineering, owing to their ability to provide real time implementable algorithms for altering  trajectories of control systems to prevent detrimental violations of input and state constraints \cite{BSK2022, GARONE2017,GK99,NicotraCM2018}. Such alterations are called for because when  control schemes are designed based on a plant model, discrepancies between the behavior of the model and the behavior of the real-world system commonly create challenges. For instance, in systems with  state constraints, parametric uncertainties in the model for which the  controller was designed may lead to undesired constraint violations, such as potentially catastrophic safety violations in aerospace systems.  Therefore, a major challenge in constrained control is the need to enforce input and state constraints in the presence of uncertain parameters.

Avoiding constraint violations in systems with parametric uncertainties has been addressed using model predictive control (or MPC) \cite{LLK22}. Proposed solutions include generating a control action which accounts for all possible trajectories of the system \cite{RACD2006}. Such methods use a local controller to confine the effects of the model uncertainties to a tube centered around a nominal trajectory \cite{D2016,PTYC2020,RKCPF2012} and to compute a sequence of one-step controllable ellipsoidal sets \cite{ACFM2008,CLK2011}.  

As detailed in \cite{GARONE2017}, reference governors provide an alternate constrained control tool that  is typically less computationally expensive than   MPC methods. Reference governors are add-on schemes that, whenever possible, preserve the response of a nominal controller designed by conventional control techniques  while ensuring that the output constraints are not violated \cite{GARONE2017}. While reference governors represent an effective tool to solve constrained control problems, few results are available on reference governors in the case of systems with parametric uncertainties \cite{CMA2000}. Most reference governor schemes exploit constraint admissible invariant sets following \cite{GT1991}\mm{,GK2002} and techniques have been developed to efficiently compute such sets for uncertain linear systems \cite{PRSD2005}. \mm{However, robust reference governors have been proposed which use trajectory prediction tubes \cite{BM1998,CAM1996,CMA2000} as well as invariance-based methods \cite{GK2002,NGK2016}.}\par
In \cite{SK2005}, a nonlinear reference governor   is used to design a load governor for fuel cell oxygen starvation protection, taking into account parametric uncertainties such as those due to imperfect control of temperature and humidity. \mm{A robust explicit reference governor scheme is employed in \cite{NNG2016} to control an unmanned aerial vehicle subject to state and input constraints.} An extended command governor is designed in \cite{BK2020} for vision-based aircraft landing on an unknown runway where the constrained linear system contains one unknown parameter. A fast reference governor for the constrained control of linear systems with parametric
uncertainties was proposed in \cite{BNK2018}.\par
This work develops a novel reference governor add-on scheme for closed-loop systems having polynomial constraints that depend on unknown constant parameters. The reference governor design uses a maximal output admissible set (or MOAS) that is computed using an extended version of the system state. When reformulated using the extended state, the constraints can be calculated using only   extreme values of the uncertain parameters as a result of the convexity property. Compared with works like \cite{BSK2022,GK99}, two significantly novel features are (a)
our state augmentation approach to transforming polynomial constraints into linear constraints and (b) 
our cancelling the uncertainties in the dynamics so the uncertainties only occur in the reformulated constraints which, due to their convexity, make it significantly easier to determine the future evolution of the constraints.

We review   mathematical preliminaries and state our problem to be addressed in 
 Section \ref{S:Prelim}. Section \ref{S:RGD} discusses the reference governor design when  uncertain polynomial constraints and input disturbances occur, and  Section \ref{S:NE} has an application to a longitudinal dynamics for a civilian aircraft. We summarize the value of our work in Section \ref{S:Con}.
\section{Preliminaries and problem formulation}
\label{S:Prelim}
We first provide notation and results from  \cite{BSK2022} and its references, followed by a statement of the problem, where the dimensions of our Euclidean spaces are arbitrary unless otherwise noted and  $\mathbb N=\{1,2,\ldots\}$.
\subsection{Kronecker products and polynomial systems base vectors}
The Kronecker product of matrices $A=[a_{ij}]\in \mathbb R^{n\times m}$ and $B=[b_{ij}]\in \mathbb R^{p\times q}$ is the $np\times mq$ matrix whose $(i,j)$ block is $a_{ij}B$ for $i=1,\ldots, n$ and $j=1,\ldots, m$, and is 
denoted by $A \otimes B$. It is non-commutative but associative and satisfies: If $A$, $B$, $C$, and $D$ are matrices of such size that one can form the matrix products $AB$ and $CD$, then
\begin{equation}\label{prodr}
(A B \otimes C D) =(A \otimes C)(B \otimes D).  
\end{equation}
Given a vector $x\in\mathbb{R}^{n_x}$ and $p\in{\mathbb{N}}$, its powers $x^{p\otimes}\in\mathbb{R}^{p n_x}$ are defined recursively using the Kronecker product:
\begin{equation}\! \! \! \begin{array}{rcl}
x^{p\otimes} \! \! &=&\! \!  \underset{i=1..p}{\otimes} x = x\otimes \left(\underset{i=1..p-1}{\otimes} x\right)\\
\! \! &=&\! \! ~x\otimes (x \otimes\ldots (x\otimes x)) \\
\! \! &=&\! \!  \left[\; \; 
x_1 x^{(p-1)\otimes}\; \; 
x_2 x^{(p-1)\otimes}\;\; 
\ldots 
x_{n_x}x^{(p-1)\otimes}
\right]^T\end{array}\! \! \! 
\end{equation}
Since the product for real numbers is commutative,  the vector $x^{p\otimes}$ contains repeated entries. This motivates introducing a vector $x^p$ obtained from $x^{p\otimes}$ by removing repetitions of entries, as follows. 
We let $x^p$  denote a vector whose entries consist of the set of  all monomials $x_1^{i_1}\ldots x_{n_x}^{i_{n_x}}$ for which $i_1+\ldots +i_{n_x}=p$.  For example, when $n_x=p=2$, we have $x=[x_1,x_2]^T$, $x^{2\otimes}=[x_1^2,x_1 x_2, x_2 x_1, x_2^2]^T$ and we can choose $x^{2}=[x_1^2,x_1 x_2, x_2^2]^T$. We fix an ordering of the entries of each vector $x^p$ throughout the sequel.
Then, and as first remarked in \cite{ChesiTAC2003}, the dimension of the vector $x^p$ is
\begin{equation}
\label{eq_sigma} 
\sigma(n_x,p)=\frac{(n_x+p-1)!}{p! (n_x-1)!}.
\end{equation}
Following \cite{ValmorbidaTAC2013}, one can recursively compute the matrices \begin{equation}\begin{array}{l}M_{c}(n_x,p)\in \mathbb{R}^{\sigma(n_x,p) \times p n_x}\; {\rm and}\\  M_{e}(n_x,p)\in \mathbb{R}^{ p n_x \times\sigma(n_x,p)}\end{array}\end{equation} that determine these relations between the power vector with and without redundant terms:
\begin{equation}
x^p = M_c(n_x,p)x^{p \otimes}\; {\rm and}\;  x^{p \otimes} = M_e(n_x,p)x^{p}.
\end{equation}
We also use the convention $x^0=1\in \mathbb R$ for any vector $x\in \mathbb R^{n_x}$.
Additionally, given $\theta\in\mathbb{R}^{n_{\theta}}$, we define the products
\begin{equation}
\label{theta_m}
    \theta x^p= \theta \otimes x^p
\end{equation}
In this case, we do not  have any repeated terms because $x^p$ and $\theta$ are different vectors. We use the notation $\theta x^p$ in the sequel. The dimension of the vector $\theta x^p$ is simply $n_{\theta}\sigma(n_x,p) $.
 Similarly, we use the notation $x^p w^q$ to denote the product $x^p\otimes w^q$ of $x^p$ by the $q$th power of a vector $w\in\mathbb{R}^{n_w}$. Again, there are no repeated terms and the dimension of $x^p w^q$ is $\sigma(n_x,p) \sigma(n_w,q)$. This product is non commutative and    $x^p w^q$ is in general different from $ w^q x^p$.
\subsection{Problem statement}
Consider a discrete time control system of the following form
\begin{equation}
\label{class_sys2}
x(k+1) = A x(k) + B v(k) + B_w w(k) , 
\end{equation}
where the state $x$ is valued in $\mathbb{R}^{n_x}$, 
the values 
$v(k) \in\mathbb{R}^{n_v}$ denote the reference governor output,  the bounded sequence $w(k)$ is valued in  $\mathbb{R}^{n_w}$
and represents an unknown disturbance input, and $A$ is a Schur matrix. Let $x_v= [x^T,v^T]^T $ denote the $\mathbb{R}^{n_x+n_v}$ valued state and reference vector.  In terms of the functions
\begin{equation}\begin{array}{l}
f_i(x_v,\theta)=  d_{i,0} \theta + \sum\limits\limits_{j=1}^p (c_{i,j}x^j_v + d_{i,j}\theta x^j_v), 
    \end{array}
\end{equation}
where the row matrices $d_{i,0}$, $c_{i,j}$, and $d_{i,j}$ are   in $\mathbb{R}^{1\times n_\theta}$, $\mathbb{R}^{1\times \sigma(n_x,j)}$ and $\mathbb{R}^{1\times n_{\theta}\sigma(n_x,j)}$ respectively, 
the system (\ref{class_sys2}) is subject to $n_c$ polynomial constraints 
\begin{equation}\label{eq_uncertain_poly_cons}\begin{array}{l}
f_i(x_v,\theta) \leq h_i\; {\rm for}\;  i\in\{1,\ldots,n_c\},\end{array}
\end{equation}
 where without  loss of generality, $h_i\geq 0$ for all $i\in\{1,\ldots,n_c\}$. The vector $\theta$ has constant but unknown bounded components. Each of them lies in a given interval and these bounds are already part of the constraints (\ref{eq_uncertain_poly_cons}). \par
In this paper, we propose a reference governor strategy to ensure that the polynomial constraints \eqref{eq_uncertain_poly_cons} of \eqref{class_sys2} are satisfied for all times $k\ge 0$ while the reference governor output $v(k)$ satisfies $v(k)\to r$ as $k\to \infty$ for a given reference vector $r$. The polynomial constraints \eqref{eq_uncertain_poly_cons} can be equivalently  written as linear constraints in $x_v,\ldots,x_v^p$ for some $p$ with uncertain coefficients that depend on $\theta$.
\section{MOAS and Reference governor design}
\label{S:RGD}

\subsection{Objectives}
Although the proposed method is valid when $r$ is constant and nonzero, we confine our discussion to the case where  $r=0$, and
where we  compute the MOAS using the choice 
\begin{equation}
\label{Eq:v}
    v_0(k+1) = \beta v_0(k)
\end{equation}
of $v$ in \eqref{class_sys2},
where the constant $\beta\in (0,1)$ 
and the initial vector $v_0(0)$ will be specified.
Later, we will use the constant $\beta$ to define the reference governor values $v(k)$ that will be implemented in the system (which, in general, will differ from the $v_0(k)$ values \eqref{Eq:v} that we will use to compute the MOAS).
The choice \rref{Eq:v} in \eqref{class_sys2}
produces a state vector
 vector $x_v=[x^\top, v^\top_0]^\top$ that evolves according to
\begin{equation}
\label{eq_xv_2}
x_v(k+1) = \Phi_{1,1} x_v(k) +  \Phi_{1,0}  w(k), \end{equation}
where \begin{equation}\Phi_{1,1} =\begin{bmatrix}A & B \\O_{n_v \times n_x} & \beta I_{n_v}\end{bmatrix}\; \; {\rm and}\; \; \Phi_{1,0}  = \begin{bmatrix}B_w \\ O_{n_v \times n_w}\end{bmatrix},\end{equation}
where $O_{r \times s}$ is the $r \times s$ matrix of all zeros.
More generally, one can  compute  matrices $\Phi_{j,i}$ such that
\begin{eqnarray}
x_v^j(k+1) = \sum\limits_{i=0}^{j} \Phi_{j,i} x_v^i(k) w^{j-i}(k),
\end{eqnarray}
using the following induction argument on $j\geq 1$.  If  this holds for $x_v^{j-1}(k+1)$ for some $j\geq 1$,  then 
we can use the relation
\begin{equation}\ashalfplus\begin{array}{rcl}
 x_v^{j\otimes}(k\! +\! 1)&=&x_v(k+1) \otimes x_v^{j-1}(k\! +\! 1)\\&=&( \Phi_{1,1} x_v(k))\otimes x_v^{j-1}(k\! +\! 1)) \\
&&+ (\Phi_{1,0} w(k))\otimes x_v^{j-1}(k\! +\! 1)
\end{array}
\end{equation}
to get 
\begin{equation}
\begin{array}{rcl}
 x_v^j(k+1) &=& 
 M_c x_v^{j\otimes}(k\! +\! 1)  \\
&=& M_c\sum\limits_{i=0}^{j-1} (\Phi_{1,1}x_v(k))\otimes 
\Phi^\sharp(i,j,k)
\\&& + M_c\sum\limits_{i=0}^{j-1} (\Phi_{1,0}w(k))\otimes 
\Phi^\sharp(i,j,k)\\
&=& \sum\limits_{i=0}^{j-1} M_{i,j} x^{i+1}_v(k)w^{j-1-i}(k)\\
&&+ \sum\limits_{i=0}^{j-1} N_{i,j} w(k)x^{i}_v(k)w^{j-1-i}(k),
\end{array}\end{equation}
where 
\begin{equation}\ashalf\begin{array}{l}
\Phi^\sharp(i,j,k)=\Phi_{j-1,i}x^i_v(k)w^{j-1-i}(k),\\
M_{i,j}=M_c\Phi_{1,1}\otimes \Phi_{j-1,i},\\
\text{and}\; N_{i,j}=M_c\Phi_{1,0}\otimes \Phi_{j-1,i}
\end{array}\end{equation}
  for all $i$, $j$, and $k$, and where we omitted the argument $(n_x+n_v,j)$ of $M_c$ and used \rref{prodr}. Hence, we can choose the coefficients $\Phi_{j,i}$ so that they satisfy
\begin{equation}\ashalf
\begin{array}{rcl}
 x_v^j(k+1) &=&
\sum\limits_{i=0}^{j-1} M_{i,j} x^{i+1}_v(k)w^{j-1-i}(k) \\ && +   \sum\limits_{i=0}^{j-1} E_{i,j}  x_v^{i}(k) w^{j-i}(k)  \\
&=& \sum\limits_{i=0}^j \Phi_{j,i} x_v^i(k) w^{j-i}(k)
\end{array}\! \! \! \! \! \! \end{equation}
where $E_{i,j}=N_{i,j}\Gamma_{i,j}$ and the $\Gamma_{i,j}$'s  are such that 
\begin{equation}\ashalfplus\! \! \! \begin{array}{l}
 w(k) \otimes x_v^i(k) w^{j-1-i}(k) = \Gamma_{i,j} x_v^{i}(k) w^{j-i}(k)\end{array}\! \! \!  \! 
\end{equation}holds for all $i$ and $j$.
Therefore, the vector \begin{equation}X_v = \begin{bmatrix}x_v & x_v^2 & \ldots & x_v^p\end{bmatrix}^T \label{Eq:Xv}\end{equation} has a discrete time dynamics of the form
\begin{equation}
\label{Eq:Ev}
    X_v(k+1) = \Phi X_v(k) +  \Phi_w X_{-1,v}(k)W(k)
\end{equation}
where
\begin{equation}\begin{array}{l}
 W = \begin{bmatrix}w & w^2 & \ldots & w^p \end{bmatrix}^T\; {\rm and}\\[.1em]
X_{-1,v} = \begin{bmatrix}1 & x_v & \ldots & x_v^{p-1}\end{bmatrix}^T.\end{array}\end{equation}
Also, by considering the special case where $w(k)$ is the zero sequence, 
$
\Phi = {\rm diag}(\Phi_{1,1}^j, j\in\{1\ldots p\})$
is a Schur matrix.

%
%
If $r$ is constant and different from $0$,  then one can perform a change of coordinates and then apply the proposed method to the error system for the difference between $x$ and the equilibrium point $\bar{x}$ defined by $\bar{x} = A \bar{x}+Br$,
where $\bar{x}$ exists and is unique since $A$ is a Schur matrix. The polynomial constraints are reformulated in the error system coordinates as well before applying the proposed method.\par
Conventional reference governors for linear systems with linear constraints use an offline computation of a finitely determined inner approximation $\tilde{O}_\infty$ to the MOAS   \cite{GARONE2017}, which can also be computed in the presence of disturbance inputs \cite{GK99}. Then the reference $v$ is updated on-line subject to the constraint $(x(k),v) \in \tilde{O}_\infty$.  In the scalar reference governor case, the computational effort is generally small. The goal of this section is to extend these ideas to linear systems subject to polynomial inequality constraints and propose procedures
\begin{itemize}
    	\item to compute  the MOAS $O_\infty$ for the system (\ref{Eq:Ev}) with the constraints (\ref{eq_uncertain_poly_cons}), and
	\item to update the reference governor based on this set.
\end{itemize}
\subsection{MOAS computation}
For \eqref{Eq:Ev} subject to the  constraints \eqref{eq_uncertain_poly_cons}, the MOAS is defined to be 
\begin{eqnarray}
\label{equ:moas1}
O_{\infty,X}=\{(x_v(0),\theta):~f_i(x_v(k),\theta) \leq h_i,\\
i=1,\ldots,n_c,~k=0,1,\ldots \}, \nonumber
\end{eqnarray}
where $x_v(k)$ is the first $n_x + n_p$ components of \eqref{Eq:Xv} at  time instant $k$ corresponding to the initial state $x_v(0)$. 
We call $O_{\infty,X}$ (and other MOASs in what follows) finitely determined if the recursive procedure for its construction in \cite{GT1991} finitely terminates (i.e., there exists $t^*$ such that  $O_{t,X}=O_{t+1,X}$ for all $t \geq t^*$ where $O_{t,X}$ is defined by
\eqref{equ:moas1} except with $k=0,1,\ldots$ replaced by $0\le k\le t$ and so
corresponds to imposing constraints on the predicted response   only for $k=0,\cdots,t$); in this case,  $O_{\infty,X}$ is representable by a finite number of inequality constraints. 

Considering that the extended state vector, denoted by $X_v$ contains the higher order powers of $x_v$  and recalling the definition of \eqref{theta_m}, it follows that the constraints of \eqref{eq_uncertain_poly_cons} can be written as a linear function of $X_v$ such that
\begin{equation}\label{newlincon}
    D_0\theta + C_0 X_v + C_1 \theta X_v \leq H,
\end{equation}
where the constant matrices $D_0$, $C_0$, and $C_1$ can be determined from   \eqref{eq_uncertain_poly_cons}, and where
$
    H = [h_i \dots h_{n_c}]^T$.
Given the linear nature of the constraints   in \eqref{newlincon}, it follows that the constraints are convex with respect to $\theta$. Thus, if  lower and upper bounds of the uncertain parameters satisfy the constraints of \eqref{eq_uncertain_poly_cons}, then so will any of the intermediate values. Computation of the MOAS then begins by   determining an initial set of constraints for every combination of the minimum and maximum values of the uncertain parameters.

Continuing  this procedure, the $n_c$ inequalities of (\ref{eq_uncertain_poly_cons}) (which depend on the unknown parameters and the extended state) are replaced by $2^{n_\theta} n_c$ inequalities which only depend linearly on the extended state.
We write them as 
\begin{equation}\label{NewConstraints}\! \! \! \begin{array}{l}
 C_{0,k} X_v \leq H_k\; {\rm for}\;   k=\{1,...,2^{n_\theta}\},\end{array}\! \! \! 
\end{equation}
for suitable constant matrices $C_{0,k}$ and $H_k$, 
by replacing the $\theta$ components from  (\ref{eq_uncertain_poly_cons}) by their lower and upper bounds to obtain the constraints \eqref{NewConstraints} where the $\theta$ values no longer appear.
Using ideas of \cite{BSK2022}, we propose a procedure to compute an inner approximation of the MOAS of our extended system; then, as in \cite{BSK2022}, a cross section (subset) of this set  is an inner approximation of the  MOAS
of \eqref{Eq:Ev} with uncertain polynomial constraints \eqref{eq_uncertain_poly_cons}. 
The inner approximation of the MOAS of the extended system is computed in two steps:

\begin{itemize}
    \item First, we assume that the MOAS of (\ref{class_sys2}) subject to the linear constraints of (\ref{eq_uncertain_poly_cons}) can be calculated and is a  robust forward invariant set that is finitely determined and compact, which implies that one has a lower and upper bound for each component of $x_v$. See the discussion between (7) and (8) in \cite{GARONE2017} of sufficient conditions for this set to be finitely determined.
    \item Second, having the lower and upper bounds of
    $x_v$ and thus of $X_v$ as well as bounds on  $w$ enables us to determine a compact set $\Omega_w$ such that the $X_{-1,v}(k)W(k)\in \Omega_w$ for all $k$. Therefore, we compute the MOAS of 
    \begin{equation}
\label{Eq:Ev2}
    X_v(k+1) = \Phi X_v(k) +  \Phi_w d(k)
\end{equation}
subject to (\ref{NewConstraints}) and the linear constraints that define the bounds on $X_v$ and under the assumption that the input disturbance satisfies $d(k)\in \Omega_w$ for all $k$. Its MOAS  can be finitely determined and is compact because $\Phi$ is a Schur matrix and each component of the extended state is bounded \cite{BSK2022}. This set is our inner approximation of the MOAS of (\ref{Eq:Ev}) subject to (\ref{NewConstraints}), because input disturbances $d(k)$ are more general than the  $X_{-1,v}(k)W(k)$ disturbance term (since the set of all disturbances defined by $X_{-1,v}(k)W(k)$ where $x_v$ follows   \eqref{eq_xv_2} and $W$ is bounded is a subset of the set of all disturbances $d$ such that $d(k)\in \Omega_w$ for all $k$).
\end{itemize}

\subsection{Reference governor update}\label{update}
Let $O_{\infty,X}$ be the MOAS \eqref{equ:moas1} considered in the previous subsection. Given an initial state $x(0)$,  the initial state $v(0)$ of the reference governor is computed so that $(x(0),v(0))\in O_{\infty,X}$, e.g., as a solution to the  optimization problem
\begin{equation} \mbox{Minimize $v^{\sf T} v$ subject to $(x(0),v) \in O_{\infty,X}$}. \end{equation}
 To reduce the computational burden for practical application of the proposed method, one can calculate  the MOAS and the initial value $v(0)$ for a grid of initial states offline. 
Then, the reference governor to be applied to \eqref{class_sys2} at times $k>0$ is
\begin{equation}
    v(k) =(1-\lambda(k)) \beta v(k-1)  
\end{equation}
where the sequence $\lambda(k)$ is valued in $[0,1]$.  The scalar variable $\lambda(k)$ can be determined using a bisection algorithm, e.g., 
\begin{equation}\label{eq-bisection}\begin{array}{l}
\lambda(k) =\max\limits_{\lambda\in[0,1]} \lambda \mbox{ subject to }\\ (x(k),\beta (1 - \lambda)  v(k-1))\in O_{\infty,X}.\end{array}
\end{equation}
\section{Numerical Example}
\label{S:NE}
\begin{figure}[!th]
\centerline{\includegraphics[width=2.8in]{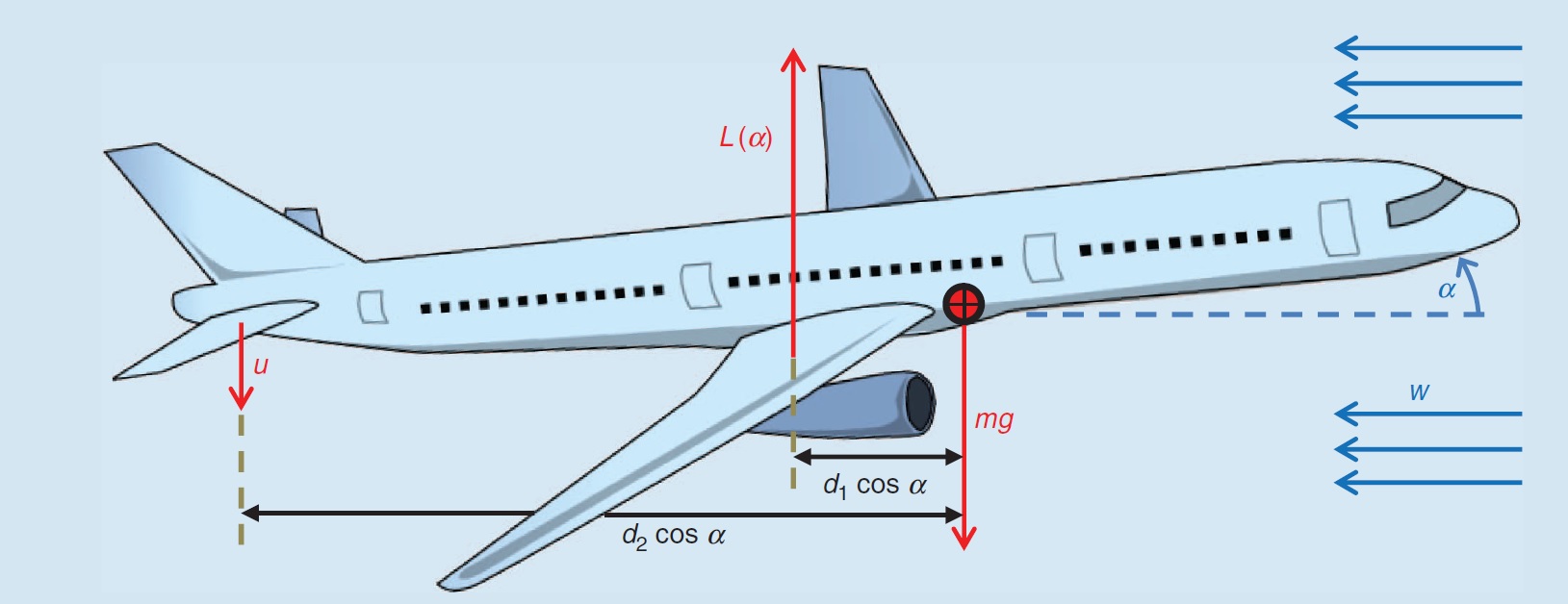}}
\caption{Longitudinal model of civilian aircraft diagram from  \cite{NicotraCM2018} showing main forces acting on the system as the weight of the vehicle mg, the lift
generated by the main wing $L(\alpha)$, and the control action $u$ generated by the tail elevator.}
\label{long}\vspace{-.5em}
\end{figure}

We consider the following aircraft longitudinal dynamics model based on \cite{NicotraCM2018, NG16} with $\cos(\alpha)$ approximated by 1, which is pictured in Fig. \ref{long}:
\begin{equation}\label{longm}
    \ddot{\alpha} = -\frac{d_1}{J}L(\alpha)+ \frac{d_2}{J} u,~~L(\alpha)=l_0+l_1 \alpha -l_3 \alpha^3,
\end{equation}
where $d_1=8m$, $d_2=40m$, $J=4.5\times10^6 Nm^2$, $l_0=2.5\times10^{5}N$, $l_1=8.6\times10^{6}N/rad$, and $l_3=4.35\times10^7 N/rad^3$.
In \eqref{longm}, 
 $J$ is the longitudinal inertia of the
aircraft, $d_1$ and  $d_2$  are the distances between
the center of mass and the two airfoils respectively, $u$ is the control
force generated by the elevator airfoil, and $L(\alpha)$ is the lift
generated by the main wing.
The angle of attack $\alpha$ is constrained by the stall limit $-0.2\times\frac{\pi}{180}\leq \alpha\leq 14.7\times\frac{\pi}{180}$rad. 
The   elevator force control input $u$  must satisfy $|u|\leq 4. 10^5$ N.\par

We apply the estimation used in  \cite[Section III]{BCMCJF2019} to estimate uncertain aerodynamic terms with $y_m=\alpha$, $\lambda=\frac{d_2}{J}$ and $F_y=-\frac{d_1}{J}L$, which provides the following formula, where $[t-T,t]$ is the window of the receding horizon strategy in \cite{BCMCJF2019}:
\begin{equation}\ashalfplus\begin{array}{rcl}
   -\frac{d_1}{J} \hat{L} (t)\! \! \! &=&\! \! \!   \frac{5!}{2T^5}\int_{t-T}^t \{[(T-\sigma)^2-4\sigma(T-\sigma)\\\! \! \! &&\! \! \! +\sigma^2]\alpha(\sigma) -\frac{d_2}{2J}\sigma^2(T-\sigma)^2 u(\sigma)\} d\sigma \label{MFC_estimation}\end{array}
\end{equation}
\mm{\begin{equation}\ashalf\begin{array}{l}
   -\frac{d_1}{J} \hat{L} (t)\\=    \frac{5!}{2T^5}\int_{t-T}^t \{[T^2-6T(t-\sigma)+6(t-\sigma)^2]\alpha(\sigma)\\  
       \; \; \; -\frac{d_2}{2J}\left[T-(t-\sigma)\right]^2(t-\sigma)^2 u(\sigma)\} d\sigma \label{MFC_estimation}\end{array}
\end{equation}}
Using $\hat{L}(\alpha)$, we apply the robust dynamic inversion control input
\begin{equation}\label{uchoice}u=-k_p(\alpha+v)-k_d \dot{\alpha}+\frac{d_1}{d_2}\hat{L}(\alpha)\end{equation} where $k_p =5.2\times10^7$ and  $k_d = 7.6\times10^6$, and where $v$ is the reference governor output.
Then, discretizing the obtained system with the sampling period $T_s=0.01s$, we obtain the (pre-stabilized) second order model \eqref{class_sys2} with $B_w=0$ and 
\begin{equation}\label{AB}
A =\begin{bmatrix}
0.9814  &  0.0072\\
-3.3347 &  0.4940 
\end{bmatrix}\; \; {\rm and}\; \;  B =\begin{bmatrix}
0.0186 \\ 3.3347
\end{bmatrix}, 
\end{equation}
where the formulas \rref{AB} were obtained by substituting \rref{uchoice} (with $\hat L$ set equal to $L$ in \rref{uchoice}) into \rref{longm}, to get a system of the form $\dot x=Fx+Gv$ with $x=[\alpha,\dot\alpha]^\top$ and then choosing
\begin{equation}\textstyle
    A=e^{F T_s}\; {\rm and}\; B=\int_0^{T_s}e^{F \ell}{\rm d}\ell G.
\end{equation}
\mm{Michael: pls note that the following formula was simplified using a change of coordinates in the integral
\begin{equation}\textstyle
    A=e^{F T_s}\; {\rm and}\; B=\int_0^{T_s}e^{F(T_s-\ell)}{\rm d}\ell G.
\end{equation}
}
This system is linear but the input inequality constraints, which are $u\leq 4 .10^5$N and $-u\leq 4 .10^5$N where $u$ is obtained by replacing $\hat{L}$ by $L=l_0+l_1 \alpha -l_2 \alpha^3$ in (\ref{uchoice}), are polynomial of order $3$. Moreover, the vector $\theta=[l_0,l_1,l_3]'$ is not precisely known; for the numerical simulations of this section, it is assumed that $\theta$ is bounded between $80\%$ and $120\%$ of the real values of the unknown parameters. In this case, the MOAS is first calculated considering only the linear constraints on $\alpha$. Using the algorithm of \cite{GT1991}, the linear MOAS is finitely determined in $t^*=75$ iterations and   defined by $105$ non-redundant linear inequalities. As this set is compact (e.g by \cite[Theorem 1]{GT1991}), we can compute bounds on all  components of the extended state and confirm that the MOAS that considers all   constraints is representable by finitely many inequalities. 

\begin{figure}[!th]\vspace{.1em}
\centerline{\includegraphics[width=3.25in]{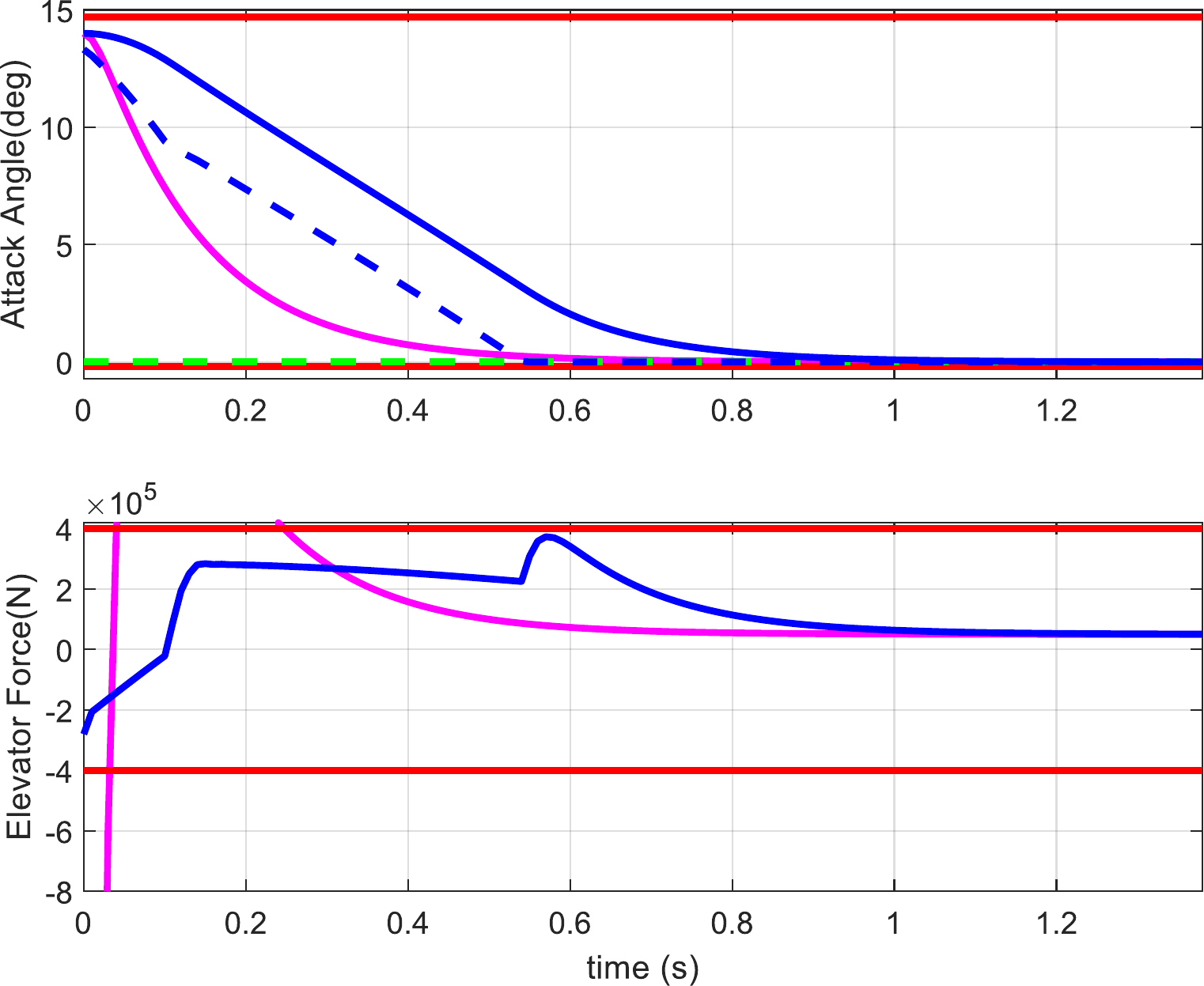}}
\caption{Constrained outputs without disturbances. Red: upper and lower limits. Magenta: when one applies the robust dynamic inversion without any reference governor, that is when $v=r=0$. Blue: when one uses the proposed reference governor. Dashed blue: evolution of the reference $v(k)$ of the reference governor. Dashed green: desired angle of attack.}
\label{fig_image_ex1}
\end{figure}

Establishing a set of constraints which consider the minimum and maximum values of all the elements of $\theta$, the robust MOAS for the entire system, including the uncertain polynomial constraints, is determined in $42$ iterations and is defined by $2062$ non-redundant linear inequalities. Figure \ref{fig_image_ex1} illustrates the constrained output responses obtained using (in blue) or not using (in magenta) the proposed reference governor when $\alpha(0)=14$ deg and $\dot\alpha(0)=0$ deg/s. In the absence of a reference governor, the control input limits are violated. However, all  output constraints are satisfied with the implementation of the proposed reference governor.
\par
\begin{figure}[!ht]\vspace{.5em}
\centerline{\includegraphics[width=3.25in]{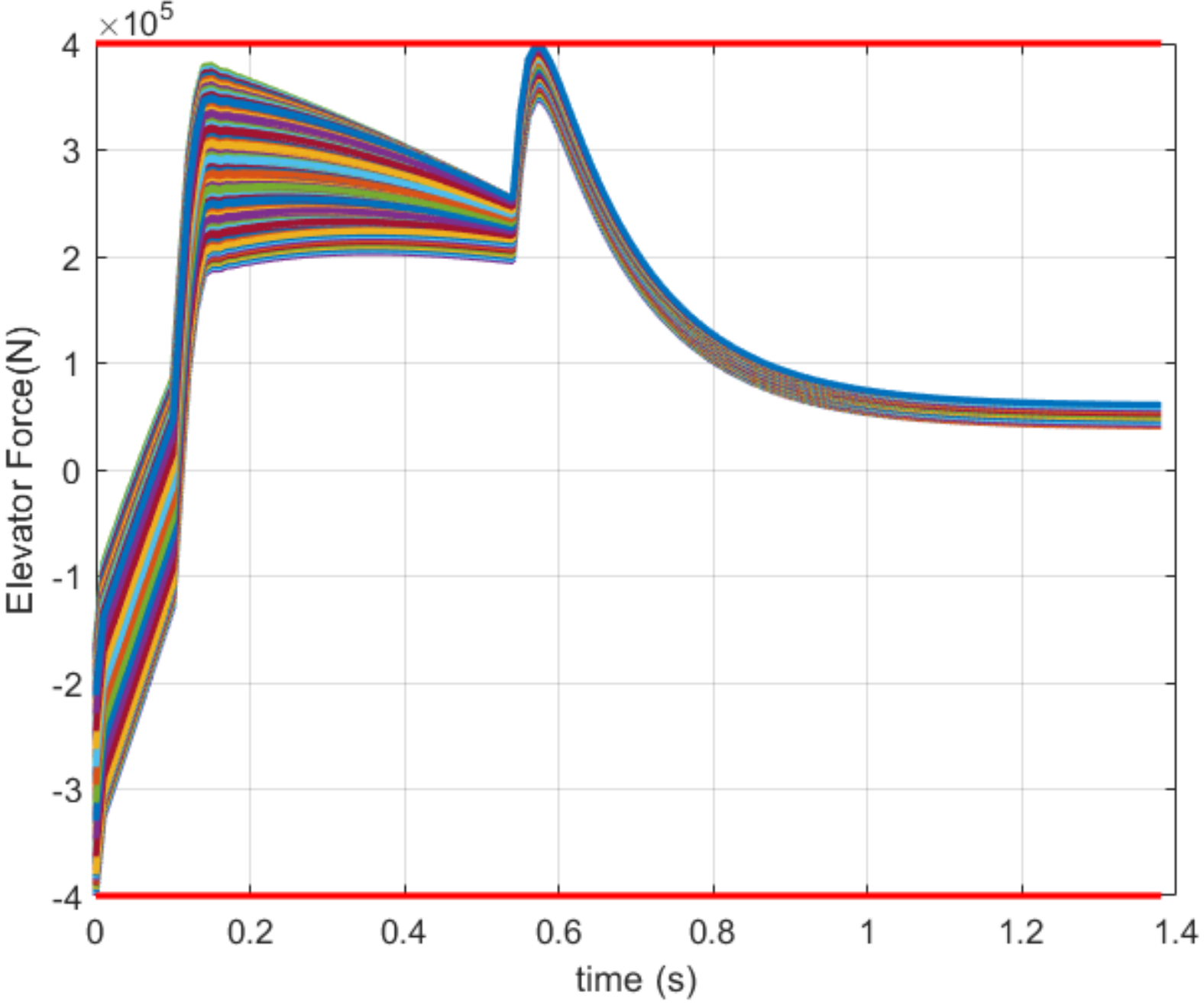}}
\caption{Constrained uncertain polynomial outputs without considering disturbances. Each color represents a different trajectory of the constrained elevator force for different values of $\theta$ within the respective bounds.}
\label{fig_image_ex1_2}\vspace{-1.5em}
\end{figure}

Figure \ref{fig_image_ex1_2} illustrates the different progressions of the constrained elevator force as the unknown parameters are varied from their minimum values to their maximum values. It is observed that regardless of the value of the bounded elements of $\theta$, or any combination thereof, the constrained output satisfies the constraints (in red).
If one considers the possibility that $\hat{L}(\alpha)$ is not a completely accurate approximation of $L(\alpha)$, the difference between the approximation and the actual value may be treated as a disturbance in the dynamics. In the following, we consider a dynamics of the form \eqref{class_sys2} where $B_w = [1~0]^T$ and $w(k)$ is bounded between $\pm 0.05$. Operating with the same conditions as the previous example, the linear MOAS is finitely determined in $t^*=47$ iterations and is defined by $81$ non-redundant linear inequalities.
\begin{figure}[!ht]
\centerline{\includegraphics[width=3.15in]{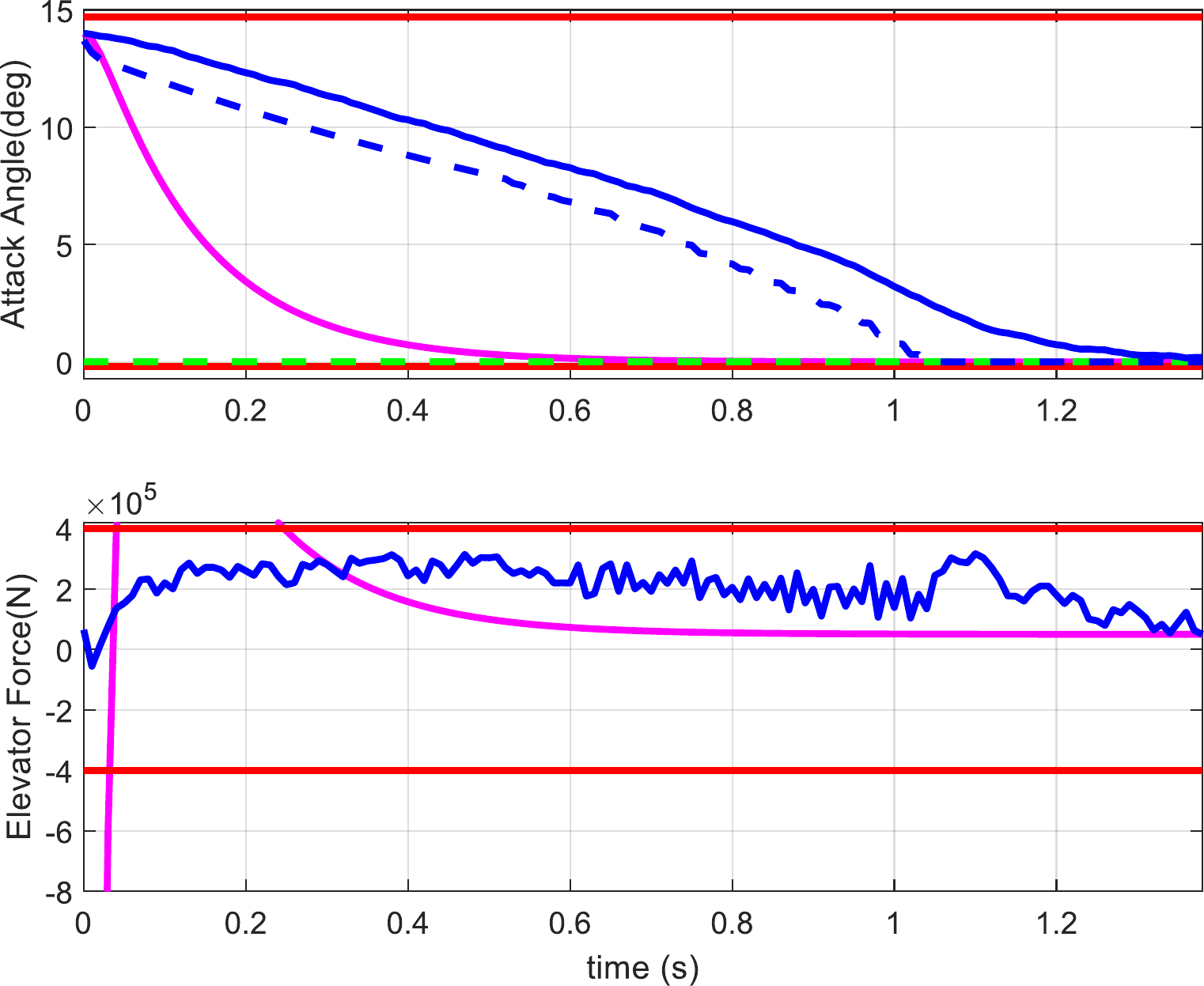}}
\caption{Constrained outputs with bounded disturbances. Red: upper and lower limits. Magenta: when one applies the robust dynamic inversion without any reference governor, that is when $v=r=0$. Blue: when one uses the proposed reference governor. Dashed blue: evolution of the reference $v(k)$ of the reference governor. Dashed green: desired angle of attack.}
\label{Fig:CO_Dist}\vspace{-.35em}
\end{figure}
\begin{figure}[!ht]\vspace{.65em}
\centerline{\includegraphics[width=3.4in]{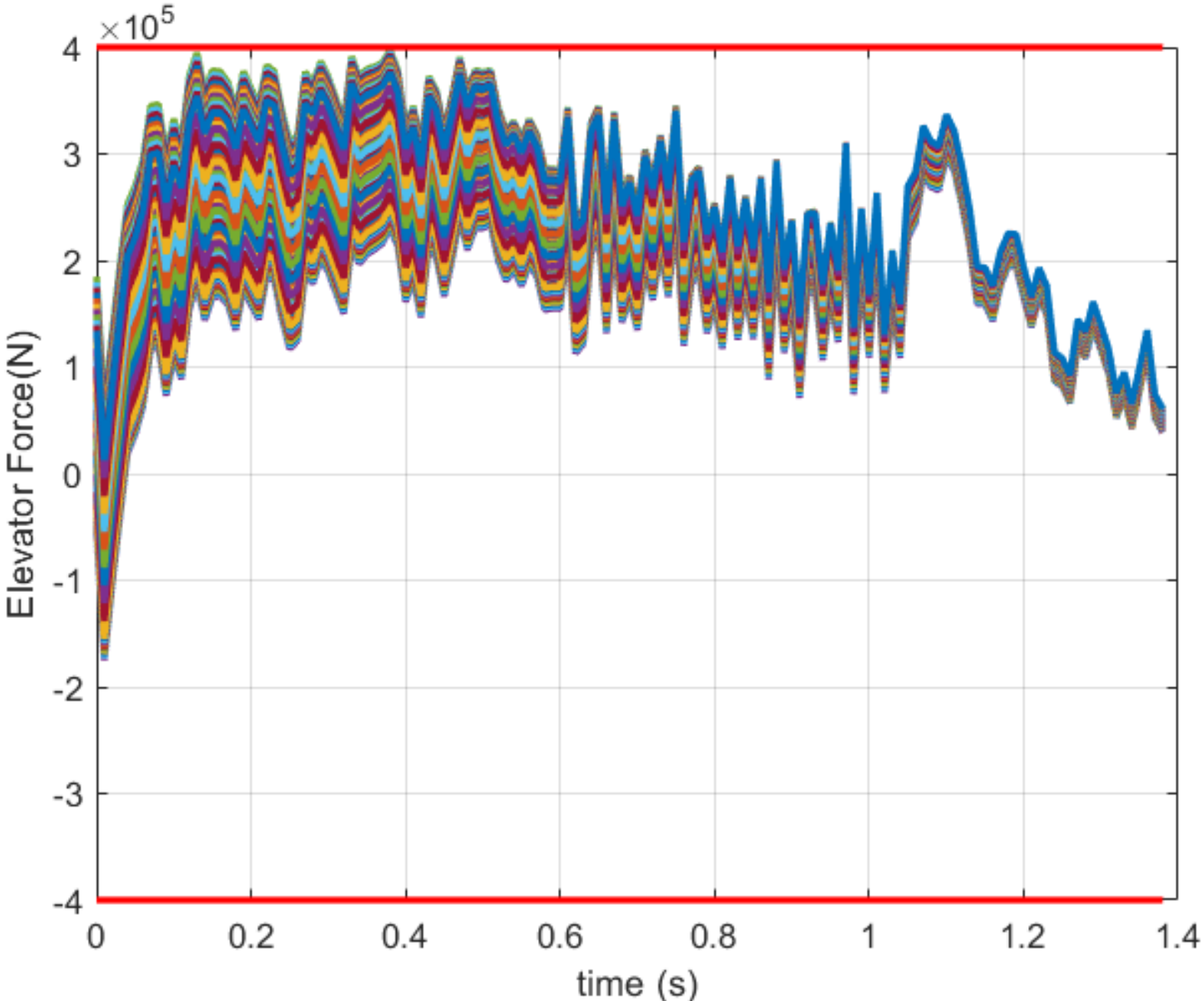}}
\caption{Constrained uncertain polynomial outputs in the presence of bounded disturbances. Each color represents a different trajectory of the constrained elevator force for different values of $\theta$ within the respective bounds.}
\label{Fig:SW_dist}\vspace{-1em}
\end{figure}
We proceed as in the previous example only now considering the bounds on the disturbances as well as the bound on the uncertain parameters. The robust MOAS for the entire system, including the uncertain polynomial constraints and disturbances, is determined in $50$ iterations and is defined by $14,067$ non-redundant linear inequalities. \par
Figure \ref{Fig:CO_Dist} illustrates the constrained outputs of the system subject to random bounded disturbances and parametric uncertainties. Constraint satisfaction is achieved although the response is slower than the case in which disturbances are not considered. Figure \ref{Fig:SW_dist} illustrates the different progressions of the constrained elevator force as the unknown parameters are varied through all possible values in the presence of this case's random disturbances. Again, it can be seen that regardless of the values of $\theta$, the constrained elevator force satisfies the constraints (in red). This illustrates the efficacy our method, and the ease with which it can be implemented. 
\section{Concluding Remarks}
\label{S:Con}

We proposed  a new reference governor add-on control scheme for systems that are subjected to uncertain polynomial constraints, using a new approach that transfers the uncertainties in the dynamics to the constraints. As in \cite{BSK2022}, the polynomial constraints were handled by extending the state and propagating the constraints through a higher dimensional LTI system. This made it possible to restate the constrained problem in terms of linear constraints on the extended state variable, and to then  project the MOAS to estimate the   MOAS in the original variables.  This projection  provides an inner approximation of the exact MOAS
of the original system.
It was shown that the constraints were convex when written in terms of the higher dimensional state. This convexity allowed a maximal output admissible set to be generated based on the minimum and maximum values of the unknown parameters. Our application to a longitudinal aircraft dynamics illustrated the efficacy of the method for systems with polynomial constraints and uncertainties, and the ease with which our novel techniques can be implemented in a significant aerospace application.

\bibliographystyle{IEEEtran}  
\bibliography{references}

\end{document}